\documentclass[floatfix]{revtex4}

\usepackage{graphicx}
\begin{document}
\bibliographystyle{revtex}


\title{Signals for Non-Commutative QED in $e \gamma$ and $\gamma \gamma$ 
Collisions}



\author{Stephen Godfrey}
\email[]{godfrey@physics.carleton.ca}
\affiliation{Ottawa-Carleton Institute for Physics 
Department of Physics, Carleton University, Ottawa, Canada K1S 5B6}

\author{M. A. Doncheski}
\email[]{mad10@psu.edu}
\affiliation{Department of Physics, Pennsylvania State University, 
Mont Alto, PA 17237 USA}


\date{\today}

\begin{abstract}
We study the effects of non-commutative QED (NCQED) in fermion pair 
production, $\gamma + \gamma \rightarrow f + \bar{f}$ and Compton scattering, 
$e + \gamma \rightarrow e + \gamma$.  Non-commutative geometries appear 
naturally in the context of string/M-theory and gives rise to 3- and 4-point 
photon vertices and to momentum dependent phase factors in QED vertices which 
will have observable effects in high energy collisions.  We consider 
$e^+ e^-$ colliders with energies appropriate to the TeV Linear Collider 
proposals and the multi-TeV CLIC project operating in $\gamma \gamma$ and 
$e\gamma$ modes.  Non-commutative scales roughly equal to the center of mass 
energy of the $e^+e^-$ collider can be probed, with the exact value depending 
on the model parameters and experimental factors.  However, we found that the 
Compton process is sensitive to $\Lambda_{NC}$ values roughly twice as large 
as those accessible to the pair production process.
\end{abstract}

\maketitle


Although string/M-theory is still developing, and the details of its 
connection to the Standard Model are still unclear, numerous ideas from 
string/M-theory have affected the phenomenology of particle physics.  The 
latest of these ideas is non-commutative quantum field theory (NCQFT) 
\cite{Douglas,ncqft}.  NCQFT arises through the quantization of strings by 
describing low energy excitations of D-branes in background EM fields.  NCQFT 
generalizes our notion of space-time, replacing the usual, commuting, 
space-time coordinates with non-commuting space-time operators.  Testable 
differences exist between QFT with commuting space-time coordinates and NCQFT.

At this time, the details of a general NCQFT model to compare to the Standard 
Model are just emerging \cite{ncsm}.  However, NCQED does exist and can be 
studied.  NCQED modifies 
QED, with the addition of a non-Lorentz invariant, momentum dependent phase 
factor to the normal $ee\gamma$ vertex, along with the addition of cubic 
($\gamma \gamma \gamma$) and quartic ($\gamma \gamma \gamma \gamma$) coupling, 
also, with non-Lorentz invariant momentum dependent phase factors.  The 
Feynman rules for NCQED are given in \cite{frncqed,hpr}.  Although the 
momentum dependent phase factors and higher dimensional operators in the 
Lagrangian arise naturally in NCQFT, the modifications, although similar, 
will in general, take on a different form than  those for 
NCQED.  We will see that the modifications of NCQFT to QED can be probed in 
$\gamma\gamma \to f\bar{f}$ and $e\gamma \to e\gamma$ collisions.  For full 
details of`< our analysis, please see Ref.~\cite{us}.

The essential idea of NCQFT is that in the non-commuting space time the 
conventional coordinates are represented by operators which no longer commute:
\begin{equation}
[\hat{X}_\mu, \hat{X}_\nu] = i\theta_{\mu\nu} \equiv {i\over 
{\Lambda_{NC}^2}} C_{\mu\nu}
\end{equation}
Here we adopt the Hewett-Petriello-Rizzo parametrization \cite{hpr} where the 
overall scale, $\Lambda_{NC}$, characterizes the threshold where 
non-commutative (NC) effects become relevant and $C_{\mu\nu}$ is a real 
antisymmetric matrix whose dimensionless elements are presumably of order 
unity.  One might expect the scale $\Lambda_{NC}$ to be of order the Planck 
scale.  However, given the possibility of large extra dimensions \cite{add,rs} 
where gravity becomes strong at scales of order a TeV, it is possible that NC 
effects could set in at a TeV.  We therefore consider the possibility that 
$\Lambda_{NC}$ may lie not too far above the TeV scale.

The $C$ matrix can be parameterized, following 
the notation of \cite{JoA}, as
\begin{equation}
C_{\mu \nu} = \left(
\begin{array}{cccc}
0 & C_{01} & C_{02} & C_{03} \\
-C_{01} & 0 & C_{12} & -C_{13} \\
-C_{02} & -C_{12} & 0 & C_{23} \\
-C_{03} & C_{13} & -C_{23} & 0 \\
\end{array}
\right)
\end{equation}
where $\sum_i |C_{0i}|^2 = 1$.  Thus, the $C_{0i}$ are related to space-time 
NC and are defined by the direction of the background {\bf E}-field.  
Likewise, the $C_{ij}$ are related to the space-space non-commutativeness and 
are defined by the direction of the background {\bf B}-field.

NCQED is beginning to attract theoretical and phenomenological interest 
\cite{hpr,ncpheno,mathews,bghh}.  Hewett, Petriello and Rizzo~\cite{hpr} have 
performed a series of phenomenological studies of NCQED at high energy, 
linear, $e^+ e^-$ colliders.  They analyzed diphoton production 
($e^+ + e^- \to \gamma + \gamma$), Bhabha scattering 
($e^+ + e^- \to e^+ + e^-$) and Moller scattering 
($e^- + e^- \to e^- + e^-$).  There are striking differences between QED and 
NCQED for all three processes; most interesting is significant structure in 
the $\phi$ angular distribution.

Mathews \cite{mathews}  and Baek, Ghosh, He and Hwang~\cite{bghh} have also 
studied NCQED at high energy $e^+ e^-$ linear colliders.  In the former case 
Mathews studied high energy Compton scattering while Baek {\it et al.,} 
studied fermion pair production in $\gamma + \gamma \to e^+ + e^-$.  
Independently of the aforementioned studies we studied Compton scattering and 
lepton pair production.  Our study uses angular distributions to enhance 
the sensitivity of measurements to $\Lambda_{NC}$, uses more 
realistic acceptance cuts, {\it etc}.


We consider linear $e^+ e^-$ colliders operating at 
$\sqrt{s} = 0.5$ and 0.8~TeV appropriate to the TESLA proposal, \cite{tesla}
$\sqrt{s} = 0.5$, 1.0 and 1.5~TeV as advocated by the NLC proponents 
\cite{nlc}, and $\sqrt{s} = 3.0$, 5.0 and 8.0~TeV being considered in CLIC 
studies \cite{clic}.  In order to estimate event rates, we assume an 
integrated luminosity of $L = 500$~fb$^{-1}$ for all cases.  We impose 
acceptance cuts on the final state particles of $10^o \leq \theta \leq 170^o$ 
and $p_{_T} > 10 \; GeV$.  Furthermore, all exclusion limits given below are 
for unpolarized electron and photon beams; the helicity structure of the 
NCQED cross section is identical to that in the SM, {\it i.e.} the 
fermion-photon couplings are vector-like, so polarization will not lead to 
an improvement in the exclusion limits.

In the pair production case, where only space-time NC enters, only one 
parameter, $\alpha$, remains in addition to $\Lambda_{NC}$.  We report 
exclusion limits for $\alpha = 0$, $\pi/4$ and $\pi/2$.  In the Compton 
scattering case, both space-space and space-time NC enter, leaving the two 
parameters, $\alpha$ and $\gamma$, in addition to $\Lambda_{NC}$.  We examine 
the two values $\gamma=0$ and $\pi/2$, and for each value of $\gamma$ give 
exclusion limits for $\alpha = 0$, $\pi/4$ and $\pi/2$.  We remind the reader 
that $\alpha$ relates to the direction of {\bf E}, whereas $\gamma$ determines 
the orientation of {\bf B}.


For the pair production process, the differential cross 
section for this process is given by:
\begin{equation}
{{d\sigma(\gamma\gamma\to f\bar{f})}\over{d\cos\theta \; d\phi}} = {{\alpha^2}\over{2s}}
\left\{ \frac{\hat{u}}{\hat{t}} + \frac{\hat{t}}{\hat{u}} - 
4 \frac{\hat{t}^2 + \hat{u}^2}{\hat{s}^2} \sin^2 
\left(\frac{k_1 \cdot \theta \cdot k_2}{2} \right) \right\}.
\end{equation}
The first two terms in the expression are the standard QED contributions, 
while the last term is due to the Feynman diagram with the cubic 
$\gamma \gamma \gamma$ coupling.  The phase factor, 
$\sin^2 \left(\frac{\mbox{$k_1 \cdot \theta \cdot k_2$}}{\mbox{$2$}} \right)$ 
only appears in this new term.  $p_1$ and $p_2$ are the momentum of the 
electron and positron, respectively, while $k_1$ and $k_2$ are the momenta of 
the incoming photons.  $\hat{s}$, $\hat{t}$ and $\hat{u}$ are the usual 
Mandelstam variables $\hat{s} = (k_1 + k_2)^2$, $\hat{t} = (k_1 - p_1)^2$ and 
$\hat{u} = (k_1 - p_2)^2$.  The bilinear product in eqn. 6 simplifies to 
\begin{equation}
\frac{1}{2} k_1 \cdot \theta \cdot k_2 = \frac{\hat{s}}{4 \Lambda_{NC}^2} 
C_{03}.
\end{equation}
%
The expression for the cross section is not Lorentz invariant due to the 
presence of the phase factor.  Note that only space-time non-commutativity 
contributes and there is no $\phi$ dependence in this case.  As 
$C_{03} = \cos \alpha$, NCQED reproduces QED for pair production when 
$\alpha = \pi/2$, and also as $\Lambda_{NC}\to \infty$.

The exclusion limits based on lepton pair production in $\gamma\gamma$ 
collisions and assuming an integrated luminosity of $L = 500 fb^{-1}$ are 
summarized in Table I of Ref.~\cite{us} for $\alpha = 0$ and $\pi/4$.  The 
values range from 220~GeV ($\sqrt{s} = 500\;GeV$) to 2.7~TeV 
($\sqrt{s} = 5.0\;TeV$). These are based on the angular distribution which, as 
already noted, gives the highest limits.  These limits could be improved by 
including three lepton generations in the final state and assuming some value 
for the lepton detection efficiency.


For the Compton scattering process, we find:
\begin{equation}
{{d\sigma(e^-\gamma\to e^-\gamma)}\over{d\cos\theta \; d\phi}} = {{\alpha^2}\over{2s}}
\left\{ -\frac{\hat{u}}{\hat{s}} - \frac{\hat{s}}{\hat{u}} + 
4 \frac{\hat{s}^2 + \hat{u}^2}{\hat{t}^2} \sin^2 
\left(\frac{k_1 \cdot \theta \cdot k_2}{2} \right) \right\}.
\end{equation}
The first two terms in the expression are the standard, QED contribution, 
while the last term is due to the Feynman diagram with the cubic 
$\gamma \gamma \gamma$ coupling.  As before, the phase factor only appears in 
this new term.

Here, $p_1$ and $k_1$ are the momenta of the initial state electron and 
photon, respectively, while $p_2$ and $k_2$ are the momenta of the final state 
electron and photon, respectively.  $\hat{s}$, $\hat{t}$ and $\hat{u}$ are the 
usual Mandelstam variables. In this case the phase factor simplifies to
\begin{equation}
\frac{1}{2} k_1 \cdot \theta \cdot k_2 = \frac{x k \sqrt{s}}{4 \Lambda_{NC}^2} 
[ (C_{01} - C_{13}) \sin \theta \cos \phi + 
(C_{02} + C_{23}) \sin \theta \sin \phi + C_{03}(1 + \cos \theta) ].
\end{equation}
Compton scattering is sensitive to both $\gamma$ and $\alpha$, so it is 
complimentary to pair production studied here.

\begin{figure}
\includegraphics[width=1.75in,angle=-90]{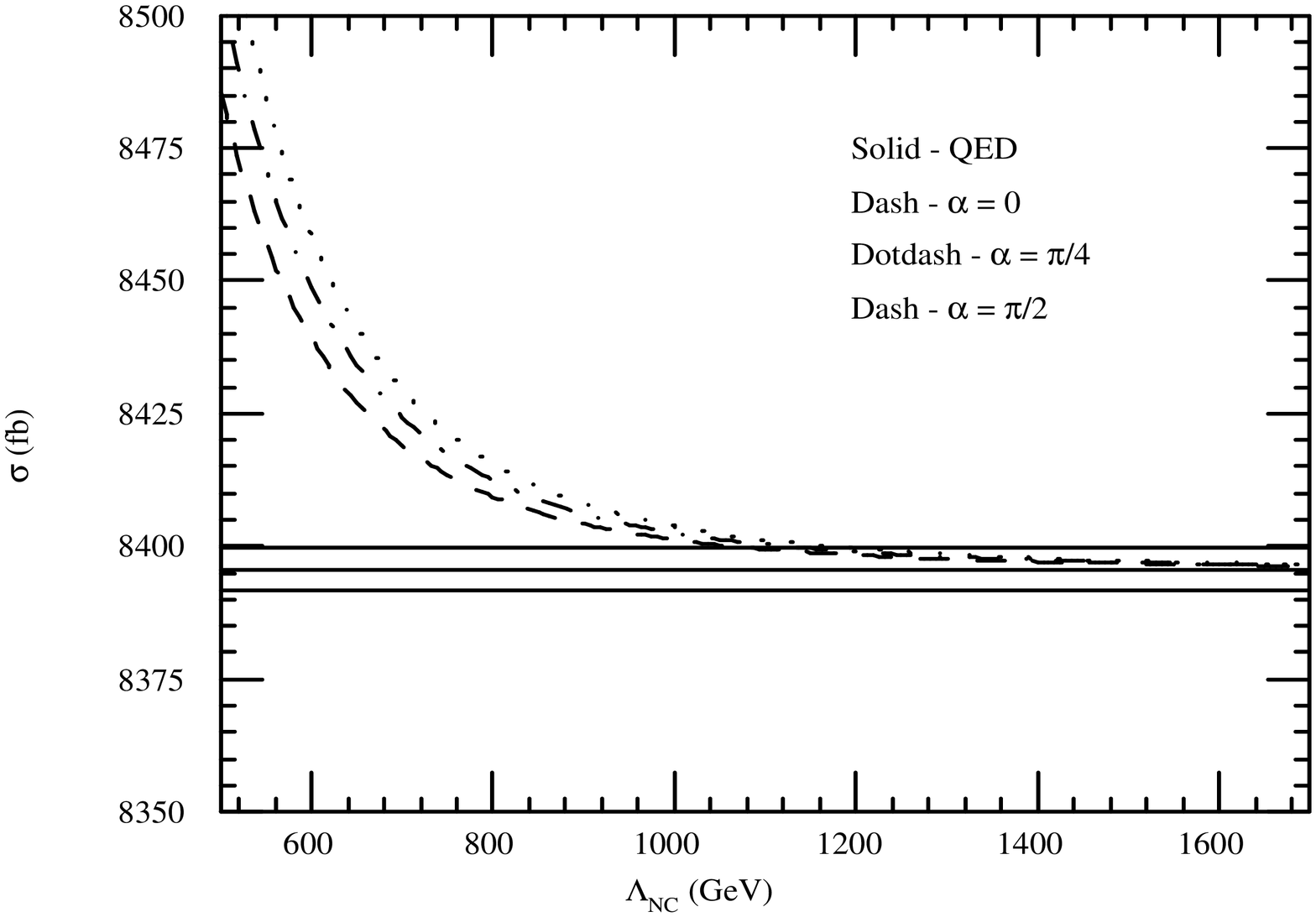}%
\includegraphics[width=1.75in,angle=-90]{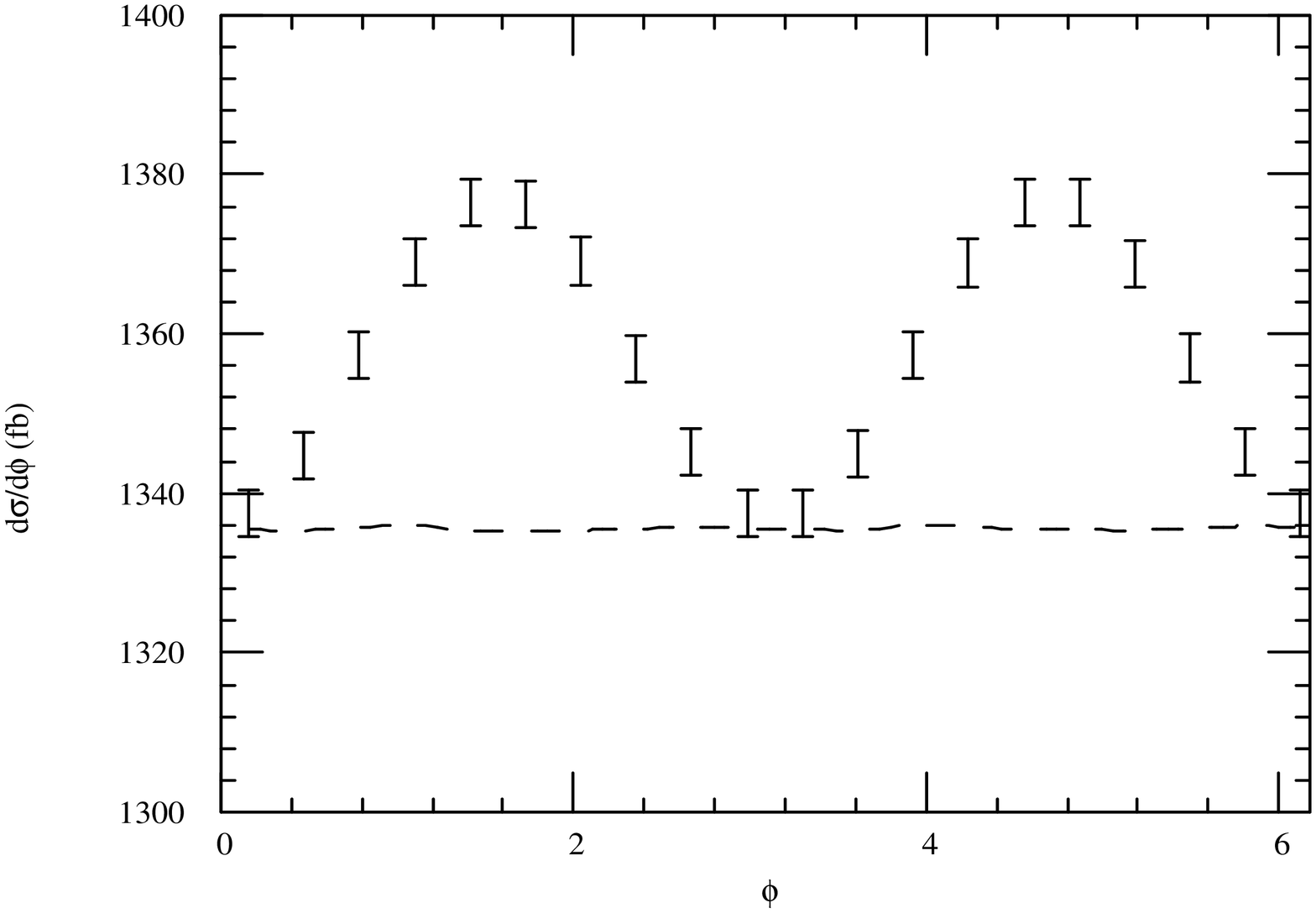}%
\caption{(a) $\sigma$ vs. $\Lambda_{NC}$ for the Compton scattering 
process with $\sqrt{s} = 500$ GeV
for $\gamma=0$.  The horizontal band represents the SM cross section 
$\pm$ 1 standard deviation (statistical) error. (b) $d\sigma/d\phi$ for the 
Compton scattering process with $\sqrt{s} = 500$ GeV and for
 $\Lambda = 500$ GeV, $\alpha = \pi/2$ and $\gamma = 0$.  The dashed curve 
corresponds to the SM angular distribution 
and the points correspond to the NCQED angular distribution including 
1 standard deviation (statistical) error.}
\label{Fig5}
\end{figure}

Fig.~\ref{Fig5}a shows the cross section $\sigma$ vs. $\Lambda_{NC}$ for QED 
and NCQED with $\alpha = 0$, $\pi/4$ and $\pi/2$, for a 
$\sqrt{s} = 0.5 \; TeV$ $e^+ e^-$ collider operating in $e \gamma$ mode.  The 
QED (solid) curve includes the central QED value and $\pm 1 \sigma$ bands 
(assuming 500~fb$^{-1}$ of integrated luminosity).  Fig.~\ref{Fig5}b shows the 
angular distribution, $d\sigma/d\phi$, for QED and NCQED with 
$\alpha = \pi/2$, and $\sqrt{s}$ = $\Lambda_{NC}$ = $500 \; GeV$.  The error 
bars in Fig.~\ref{Fig5}b assume 500~fb$^{-1}$ of integrated luminosity.  Note 
that there is no $\phi$ dependence for $\alpha=0$ since for this case both 
{\bf E} and {\bf B} are parallel to the beam direction.  In contrast, when 
$\alpha=\pi/2$, {\bf E} is perpendicular to the beam direction which is 
reflected in the strong oscillatory behavior in the $\phi$ distribution.  The 
exclusion limits obtainable from Compton scattering are summarized in Table II 
of Ref.~\cite{us} for $L = 500 fb^{-1}$, with and without a 2\% systematic 
error.  The exclusion limits range from 545~GeV ($\sqrt{s} = 500\;GeV$) to 
7.4~TeV ($\sqrt{s} = 5.0\;TeV$).

The pair production process is only sensitive to space-time NC and is 
therefore insensitive to $\gamma$.  As $\alpha$ increases towards $\pi/2$ the 
deviations from SM decrease towards zero, with $\alpha = \pi/2$ being 
identical to the SM.  On the other hand, the Compton scattering process is 
sensitive to both space-space and space-time NC as parametrized by $\gamma$ 
and $\alpha$.  On the whole, we found that the Compton scattering process is 
superior to lepton pair production in probing NCQED.  Despite significantly 
smaller statistics, the large modification of angular distributions 
(see Fig.~\ref{Fig5}b) leads to higher exclusion limits, well 
in excess of the center of mass energy for all colliders considered.

\end{document}